\documentclass[journal]{IEEEtran}
\usepackage{amsmath,amsfonts}
\usepackage{algorithmic}
\usepackage{algorithm}
\usepackage{array}
\usepackage[caption=false,font=normalsize,labelfont=sf,textfont=sf]{subfig}
\usepackage{textcomp}
\usepackage{stfloats}
\usepackage[hyphens]{url}
\usepackage{hyperref}
\usepackage[hyphenbreaks]{breakurl}
\usepackage{verbatim}
\usepackage{graphicx}
\usepackage{tikz}
\usepackage{cite}
\begin{document}

\title{LiDAR for Robust Exoatmospheric Positioning}

\author{L.M. Arthur$^*$, R.S. Kemp$^*$
\thanks{$^*$Laboratory for Nuclear Security and Policy, Department of Nuclear Science and Engineering, Massachusetts Institute of Technology}%
}

\markboth{Typeset in the IEEE style}%
{L.M. Arthur, R.S. Kemp: Intensity Modulated Direct Detection LiDAR for Robust Exoatmospheric Positioning}

\maketitle

\begin{abstract}
We present an active method for robust exoatmospheric positioning using space-based intensity modulated direct detection (IMDD) LiDAR with a spacecraft-based transmitter and receiver and a constellation of orbital reflectors. We discuss the precision of the method, the power requirements, the advantages of such a method over existing GPS/GNSS systems, and potential applications.
\end{abstract}

\begin{IEEEkeywords}
LiDAR, GNSS, positioning, guidance
\end{IEEEkeywords}

\section{Introduction}
\IEEEPARstart{M}{any} spacecraft require position and velocity information to plan and execute maneuvers for orbital insertion, rendezvous, or reentry. While early space missions did not require precision, modern applications such as satellite servicing and maintenance, the use of reusable launch vehicles, intercontinental ballistic missile guidance and intercept, and some satellite-to-satellite communications, require precise positioning and velocity information.

Global Navigation Satellite Systems (GNSS), such as the United States' Global Positioning System (GPS), can be used for precise positioning both on the Earth's surface and in low Earth orbit (LEO).\cite{gps} However, current GNSS systems use a small number of large, complex, and expensive satellites, which cannot be repaired or promptly replaced, meaning that disabling only a few satellites could disrupt the system over a large area. The low received power and the long distances involved also mean that GNSS is susceptible to signal spoofing and jamming.\cite{bbcRussiaJamming} In the face of proliferating anti-satellite weapons and electronic warfare systems, it may be a priority for governments and commercial entities to seek an alternative method of space navigation that is more robust against interference from adversaries.

An existing alternative to GNSS is the use of ground-based tracking. However, radar and optical signals are subject to atmospheric distortions, degrading positional accuracy. Persistent tracking with extended integration times can overcome atmospheric distortions, but this is not suitable for guiding short time-scale maneuvers. Additionally, the limited view of a single ground station means that a large network would be required for persistent tracking throughout an orbit or trajectory, and tracking may be unavailable over contested or remote regions of Earth. Ground tracking is also subject to adversarial disruption. Ground data must be aggregated from a distributed network of stations and transmitted promptly to the vehicle, during which time it may be subject to jamming, spoofing, or other interference.

We introduce a more robust approach to space navigation that uses autonomous multilateration against positional fiducials, or benchmarks in the language of geodesy. This avoids trans-atmospheric signals and does not depend on jammable GNSS satellites. Distances to benchmarks can be determined by time-of-flight measurements. In space, however, a benchmark's position is time dependent as it orbits the Earth. Correspondingly, a benchmark's location must be estimated using its orbital ephemerides and a precise clock. The exoatmospheric approach is resistant to disruption because the position measuring system is space-based, self-contained, and can take advantage of a large number of brightly reflecting orbiting objects as benchmarks. 

\section{LiDAR for Distance Ranging}
In principle, RADAR, LiDAR, or any other time-domain reflectometry can be used to estimate the range to a benchmark. We illustrate the proposed method using Intensity Modulated Direct Detection (IMDD) LiDAR. We judge that a LiDAR-based system is likely to be more robust to interference from off-axis (sidelobe) signals. LiDAR hardware has already been demonstrated in space for applications such as altimetry and remote sensing of planetary surfaces, and has been used for relative positioning of spacecraft for rendezvous and docking.\cite{Mcmanamon2019, Christian2013} Long-range lasers have also been used for optical communications in space, such as in the ongoing Deep Space Optical Communications (DSOC) project, and have even been demonstrated at interplanetary ranges.\cite{lasercomms} The range to a reflecting benchmark can be determined using the signal time-of-flight, which, when unimpeded by atmospheric effects, is simply:
\begin{equation}
    R = \frac{c \Delta t}{2}
\end{equation}

If four or more reflector satellites are within the vehicle's field of view and transmitter range, multilateration can provide a fully-constrained position measurement to update the navigation software's state estimate for the vehicle. The positions of the reflectors must be known to the vehicle in advance. That can be achieved by pre-loading a set of orbital models for all well-tracked brightly reflecting benchmark objects (see Section~\ref{sec:reflectors}) into the vehicle's guidance computer. This dataset must be established in advance and maintained using ground-based tracking systems. However, unlike ground-based tracking of a maneuvering vehicle, only a small number of ground stations are required for determining orbital ephemerides. These ground-based stations can be located in protected territory, and no ground-to-space communication is needed. 

Intensity modulated direct detection (IMDD) is a LiDAR approach that allows for the precise measurement of signal time of flight with a pseudo-random noise (PRN) sequence and without requiring a frequency modulated laser or an optical local oscillator to detect the frequency modulations. A photodetector could register the intensity of the returned signal, and the time of flight could be determined by cross-correlating the received signal with the transmitted signal. This approach allows for the disambiguation of multiple returns. Optical intensity modulation at frequencies above 1~GHz and transmitted powers above 100~mW have been demonstrated.\cite{Cox1997} As we show below, this could be used to achieve nanosecond precision and provides sufficient power for the ranges over which the laser might operate.

\subsection{Accuracy and Error Analysis}
The error in the estimated position of the vehicle depends on the uncertainty in the measured round trip time-of-flight to the reflectors, and on the accuracy with which the positions of the reflectors are known. The accuracy of the estimated position of the reflectors depends, in turn, on the error in the absolute timekeeping and the accuracy of the ephemerides of the reflectors in the reflector database.

The PRN signal modulated at gigahertz frequencies allows for time-of-flight measurements with nanosecond precision from point-source reflectors.\cite{Cox1997, ilrs} This contribution to the vehicle's position uncertainty scales linearly with the time of flight uncertainty. A 1~ns uncertainty in the time of flight corresponds to a 0.15~m uncertainty in the position from a single measurement. For $n$ pulses, the range uncertainty $\sigma_n$ can be related to the single-pulse uncertainty $\sigma_{sp}$ by the square root of the number of pulses.
\begin{equation}
    \sigma_{n} \approx \frac{\sigma_{sp}}{\sqrt{n}}
\end{equation}
For a one hundred bit PRN, up to $10^7$ measurements could be made per second. Only $10^5$ measurements are needed to suppress the aleatoric uncertainty to the millimeter scale.

Reflected signals are convolutions of the transmitted signature and the reflector geometry. For well-characterized reflectors, the convolution can be known in advance, and the centroid can be identified to millimeter accuracy, as demonstrated by the International Laser Ranging Service (ILRS) for geodesy reflectors.\cite{ilrs} Reflectors for which deconvolution is not possible produce degraded range measurements, with errors on the scale of the radius of the reflector.

The locations of the reflector satellites are estimated from pre-loaded ephemerides evaluated during flight. Use of the ephemerides database requires the guidance system to maintain absolute timekeeping, calibrated to the time standard used by the ground tracking system that established the corresponding ephemeris. The onboard time standard could be calibrated to the ground tracking system prior to launch with 10~ns accuracy, matching GPS timekeeping standards.\cite{gps} The standard could then be maintained during the duration of the flight with a small atomic clock for the absolute time and a local oscillator for measuring time differentials. For the time error contribution to reflector position uncertainty to be below 1~cm for reflector satellites at 1000~km altitude, moving along circular orbits at approximately 7.5~km/s, the guidance system's absolute time estimate must be accurate to one microsecond. Commercially available chip-scale atomic clocks (CSACs) have linear fractional frequency drifts of $ 2 \times 10^{-8}$~days$^{-1}$, with Allan deviations of $2.5 \times 10^{-11}$ at integration times of 250~s, at 75~mW power.\cite{Knappe2004, Newman2019} Commercially available CSACs rated for spaceflight are larger, but have Allan deviations of $ 1 \times 10^{-11}$ at 1000~s, and linear drifts of $ 1 \times 10^{-11}$~days$^{-1}$, and 120~mW power.\cite{microsemiCSAC} NASA's Deep Space Atomic Clock (DSAC), demonstrated in 2019, required a 14-liter volume and a 44~W power supply, but has a Allan deviation of $3 \times 10^{-15}$ at one day, with linear fractional frequency drift of $3 \times 10^{-16}$~days$^{-1}$, meeting the microsecond accuracy requirement for up to 10 years after launch.\cite{Seubert2022} Based on these performance specifications, we conservatively estimate the total reflector satellite position uncertainty as below 0.1~m.

\begin{equation}
    \sigma_{ref} = \sqrt{\sigma_{eph}^2 + (\sigma_t v_{orb})^2}
\end{equation}

The position uncertainty $\sigma_{pos}$ for the vehicle along a given measurement axis can then be determined by summing in quadrature the reflector position uncertainty $\sigma_{ref}$ and the range uncertainty $\sigma_n$.
\begin{equation}
    \sigma_{pos} \approx \sqrt{\sigma_{ref}^2 + \sigma_n^2}
\end{equation}
Because we expect $\sigma_n\ll\sigma_{ref}$, the overall accuracy of the position estimate would be determined by the quality of the ephemerides and the on-board clock. If multiple reflectors are within view, errors from ephemerides estimation can be further reduced. Overall positional uncertainties should be in the centimeter to decimeter range, depending on the elapsed time since the calibration of the clock. 

\section{Reflector Infrastructure and Orbital Models}
\label{sec:reflectors}
Positional benchmarks could in principle be any brightly reflecting object in orbit. However, purpose-built reflecting satellites have the advantage of known geometry, which allows deconvolution of the reflected signal for greater positional accuracy. The reflectors themselves can be fabricated from small, lightweight, corner cubes, which have extremely high optical cross-sections. These reflectors can be tiled on dedicated spherical satellites, or piggybacked on other satellites which have different primary missions.

There is a trade off between placing the reflector infrastructure in LEO or MEO. Positioning reflectors in LEO would reduce the required signal strength, but necessitate more satellites and ground stations for global coverage. Lower orbital altitudes would also increase the drag on the reflectors, requiring more frequent replacement of the constellation and more persistent tracking of the reflectors. Additionally, errors arising from orbital perturbations from higher order modes in the gravitational field, and pressure from radiation reflected by the Earth, decrease with increasing orbital altitude. Placing the reflectors higher, therefore, such as in MEO, reduces these drawbacks but at the cost of requiring higher signal strengths, which scale as distance to the fourth power. Finally, placing the reflectors at higher altitudes may also make them less susceptible to kinetic attack or interception. A detailed accounting of power requirements for varying reflector configurations is given in section~\ref{subsection:power_requirements}.

The International Laser Ranging Service continuously tracks dedicated reflectors in MEO with millimeter precise measurements for geodesy and other research purposes.\cite{ilrs}. These measurements from a global network of ground stations are consolidated into ephemerides data products with position data reported to millimeter precision. Because of the extent of historical data, collected since 1976 in the case of LAGEOS, orbits are well-characterized. Orbital predictions can be accurate to millimeters over a seven-day prediction window, and are updated several times per day. Those orbital predictions can be interpolated with Lagrange polynomials to efficiently determine each reflecting satellite's state vector at any given time, with accuracy in the tens of millimeters.\cite{ilrspred}

\section{Hardware Requirements}
To approach the precision of GNSS systems, the time-of-flight positioning system would need meter-scale precision outside of the atmosphere. For an optical signal, this would require nanosecond timing capabilities for the transmission and reception of signals.

\subsection{Power Requirements}
\label{subsection:power_requirements}
For a diffraction-limited LIDAR system, the received power $P_r$ is related to the transmitted power $P_t$ by the the optical cross-section $\sigma$ of the reflector, the effective collecting area of the receiver $A_e$, the distance $R$ between the vehicle and reflector, the wavelength of the signal $\lambda$, and the beam waist of the transmitted signal $w_0$.
\begin{equation}
  \label{equation:range}
    P_r = \frac{4 P_t w_0^2 \sigma A_e}{\pi^2 R^4 \lambda^2} 
\end{equation}

To determine the received power required to exceed a given signal-to-noise ratio, the expected noise power must be estimated. The dominant noise source the system must tolerate is radiation flux from the background sky.\cite{Mcmanamon2019, Hemmati2011} The constant background can be subtracted from the received power, however, the background radiation arriving at the detector is Poissonian, with expected variance equal to the background power. Assuming a background flux equivalent to that of a reference magnitude star in the visual band, $m_v = 0$, the background flux would be approximately $10^{-23}$~W~m$^{-2}$~Hz$^{-1}$, over the 88~nm width of the visual band, corresponding to a background of $10^{-9}$~W~m$^{-2}$. A commercially available, nanometer-width bandpass filter could further suppress the noise background to $10^{-11}$~W~m$^{-2}$. This constant background can be subtracted away, leaving Poissonian fluctuations of $10^{-16}$~W, for a 0.01~m$^2$ receiver. To overcome the background noise and achieve a signal-to-noise ratio of 10, the received power should then be greater than $10^{-15}$~W.

Following these relationships, the approximate power requirements for a 550~nm laser system can be estimated for a set of transmission ranges and reflector cross-section. Purpose-made retro-reflectors, such as the LAGEOS and Etalon satellites, have measured optical and IR cross sections of $10^7$~m$^2$.\cite{ilrs_cross_sections} In contrast, a non-purpose-made reflector might have an $\mathcal{O}(1)$~m$^2$ cross section. Assuming a beam waist of 1~cm, and a receiving area of 0.01~m$^2$, the power requirements for different distances, corresponding approximately to LEO-to-MEO (5000~km) and LEO-to-LEO (1000~km), are shown in Table~\ref{tab:power}.

\begin{table}
    \centering
    \begin{tabular}{|c|c|c|}
        \hline
        Range (km) & $\sigma$ (m$^2$) & $P_t$ (W) \\
        \hline
        5000 & $1$ & $5 \times 10^{5}$ \\
        5000 & $10^7$ & $5 \times 10^{-2}$ \\
        1000 & $1$ & $7 \times 10^2$ \\
        1000 & $10^7$ & $7 \times 10^{-5}$ \\
        \hline
    \end{tabular}
    \vspace{1mm}
    \caption{Power requirements for a 550~nm LiDAR system with a 0.01~m beam waist and a 0.01~m$^2$ receiver area with a $10^6$~m$^2$ optical cross section, for a signal-to-noise ratio of 10, and a background of $10^{-9}$~W~m$^{-2}$~Hz$^{-1}$ across the visual band.}\label{tab:power}
\end{table}

For the system to be powered for a flight time of up to one hour by a battery comparable to a standard car battery, approximately 1~kWh, assuming a 10\% laser efficiency, the transmitted power should be less than 100~W. Table~\ref{tab:power}, makes clear that such is feasible but only for nearby reflectors or purpose-made reflectors. If the system only needs to operate for $\sim$10~seconds, which may be adequate for a single course correction to inertial navigation, the required power could be as high as $10^4$~W.

\subsection{Beam Steering}
To ensure that the transmitted beam reaches the targeted reflector, the uncertainty in the vehicle orientation and the beam steering would need to be less than the beam width of the transmitted signal. For a diffraction-limited beam, the beam width of the first Airy disk can be estimated as 
\begin{equation}
    \theta \approx 1.22 \frac{\lambda}{w_0}
\end{equation}
where $\lambda$ is the wavelength of the signal and $w_0$ is the beam waist. For a 550~nm wavelength and a 1~cm beam waist, the beam width would be approximately 10$^{-4}$ radians, within the performance of existing inertial guidance systems over timescales of ten days or less.\cite{Titterton2004} Beam-steering could be accomplished through the use of commercially available mechanical gimbals coupled to fast steering mirrors, which have the micro-radian angular resolution and millisecond response times required for deep-space optical communication.

\section{Conclusion}
A diffraction-limited, IMDD LiDAR system could be used for exoatmospheric positioning with decimeter accuracy on a low power-budget. With monthly clock calibrations and periodic updates to the reflector satellite ephemerides, position measurements could be accurate to 10--20~mm. Without such updates, accuracy would degrade over the course of weeks, while bias drift in the inertial guidance system used to orient the transmitter would interfere with beam-steering after ten days without calibration. With calibration of the beam-steering system, we estimate that accuracy better than 1~m would be possible for up to 100 days, making the system robust to disruption in contested environments.

Such a system could provide global coverage with a constellation of reflector satellites, which would be resilient to attack or interference, meaning that the system would be able to operate in contested environments. Because the reflectors would be compact and inert, the constellation could be augmented or replaced promptly and economically. The narrow optical beams would limit side-lobe interference, and tuning the optical frequency to attenuate in the atmosphere would further protect against interception, jamming, or spoofing from the ground. This robust system could support a variety of space missions requiring precise position control, including satellite servicing, satellite-to-satellite communication, precision-guided munitions, and reusable launch platform recovery.

\section*{Acknowledgments}
This research was supported in part by the Carnegie Corporation of New York.

\section*{Conflicts of Interest}
The authors are employees of the Massachusetts Institute of Technology, which has filed a pending United States patent application for exoatmospheric positioning, based on the work presented in this preprint. 

\bibliographystyle{IEEEtran}
\bibliography{exoatmospheric_positioning}

\begin{IEEEbiographynophoto}{L.M. Arthur} is a Technical Associate in the Laboratory for Nuclear Security and Policy in the Department of Nuclear Science and Engineering at the Massachusetts Institute of Technology.
\end{IEEEbiographynophoto}

\begin{IEEEbiographynophoto}{R.S. Kemp} is Associate Professor of Nuclear Science and Engineering at the Massachusetts Institute of Technology and Director of the Laboratory for Nuclear Security and Policy.
\end{IEEEbiographynophoto}

\vfill

\end{document}